\documentclass[aps,prl,groupedaddress,fleqn,twocolumn]{revtex4}
\pdfoutput=1
\usepackage{graphicx}
\usepackage{amsmath}
\usepackage{gensymb}

\begin{document}
\title{Direct links between dynamical, thermodynamic and structural properties of liquids: modelling results}
\author{L. Wang$^{1}$}
\author{C. Yang$^{1}$}
\author{M. T. Dove$^{1}$}
\author{Yu. D. Fomin$^{2}$}
\author{V. V. Brazhkin$^{2}$}
\author{K. Trachenko$^{1}$}
\address{$^1$ School of Physics and Astronomy, Queen Mary University of London, Mile End Road, London, E1 4NS, UK}
\address{$^2$ Institute for High Pressure Physics, RAS, 142190, Moscow, Russia}


\begin{abstract}
We develop an approach to liquid thermodynamics based on collective modes. We perform extensive molecular dynamics simulations of noble, molecular and metallic liquids and provide the direct evidence that liquid energy and specific heat are well-described by the temperature dependence of the Frenkel (hopping) frequency. The agreement between predicted and calculated thermodynamic properties is seen in the notably wide range of temperature spanning tens of thousands of Kelvin. The range includes both subcritical liquids and supercritical fluids. We discuss the structural crossover and inter-relationships between structure, dynamics and thermodynamics of liquids and supercritical fluids.
\end{abstract}

\maketitle

\section{Introduction}

It is an interesting fact that the liquid state has proven to be difficult to describe by theory throughout the history of condensed matter research \cite{frenkel,boonyip,march,march1,enskog,baluca,ziman,zwanzig,hansen1,faber,hansen2,landau}. The problem extends beyond condensed matter and exists in other areas where strong interactions are combined with dynamical disorder such as field theory.

In a weakly-interacting system such as a dense gas, the potential energy is much smaller than the kinetic energy. These systems are amenable to perturbation treatment giving corrections to the non-interacting case \cite{enskog}. Perturbation approaches have been widely explored to calculate liquid thermodynamic properties but have not been able to agree with experiments. For example, the analysis of tractable models such as van der Waals or hard-spheres systems returns the gas-like result for the liquid constant-volume specific heat $c_v=\frac{3}{2}k_{\rm B}$ \cite{landau,pre,prl}. This is in contrast to experimental results showing that $c_v$ of monatomic liquids close to the melting point is nearly identical to the solid-like result, $c_v=3k_{\rm B}$ and decreases to about $2k_{\rm B}$ at high temperature \cite{grimvall,wallace}. As expected on general grounds, the perturbation approach does not work for strongly-interacting systems.

Strong interactions are successfully treated in solids, crystals or glasses, where the harmonic model is a good starting point and gives the most of the vibrational energy. However, this approach requires fixed reference points around which the energy expansion can be made. With small vibrations around mean atomic positions, solids meet this requirement but liquids seemingly do not: liquid ability to flow implies that the reference lattice is non-existent.

Therefore, liquids seemingly have no simplifying features such as small interactions of gases or small displacements of solids \cite{landau}. In other words, liquids have no small parameter. One might adopt a general approach not relying on approximations and seek to directly calculate the liquid energy for a model system where interactions and structure are known. This meets another challenge: because the interactions are both strong and system-dependent, the resulting energy and other thermodynamic functions will also be strongly system-dependent, precluding their calculation in general form and understanding using basic principles, in contrast to solids and gases \cite{landau}. Consistent with this somewhat pessimistic view, the discussion of liquid thermodynamic properties has remained scarce. Indeed, physics textbooks have very little, if anything, to say about liquid specific heat, including textbooks dedicated to liquids \cite{frenkel,boonyip,march,march1,enskog,baluca,ziman,zwanzig,hansen1,faber,hansen2,landau}.

As recently reviewed \cite{ropp}, emerging evidence advances our understanding of the thermodynamics of the liquid state. The start point is the early theoretical idea of J Frenkel \cite{frenkel} who proposed that liquids can be considered as solids at times smaller than liquid relaxation time, $\tau$, the average time between two particle rearrangements at one point in space. This implies that phonons in liquids will be similar to those in solids for frequencies above the Frenkel frequency $\omega_{\rm F}$:

\begin{equation}
\omega>\omega_{\rm F}=\frac{1}{\tau}
\label{omega}
\end{equation}

The above argument predicts that liquids are capable of supporting shear modes, the property hitherto attributable to solids only, but only for frequencies above $\omega_{\rm F}$.

We note that low-frequency modes in liquids, sound waves, are well-understood in the hydrodynamic regime $\omega\tau<1$ \cite{hydro}, however Eq. (1) denotes a distinct, solid-like elastic regime of wave propagation where $\omega\tau>1$. In essence, this suggests the existence of a cutoff frequency $\omega_{\rm F}$ above which particles in the liquid can be described by the same equations of motion as in, for example, solid glass. Therefore, liquid collective modes include both longitudinal and transverse modes with frequency above $\omega_{\rm F}$ in the solid-like elastic regime and one longitudinal hydrodynamic mode with frequency below $\omega_{\rm F}$ (shear mode is non-propagating below frequency $\omega_{\rm F}$ as discussed below).

Recall the earlier textbook assertion \cite{landau} that a general thermodynamic theory of liquids can not be developed because liquids have no small parameter. How is this fundamental problem addressed here? According to Frenkel's idea, liquids behave like solids with small oscillating particle displacements serving as a small parameter. Large-amplitude diffusive particle jumps continue to play an important role, but do not destroy the existence of the small parameter. Instead, the jumps serve to modify the phonon spectrum: their frequency, $\omega_{\rm F}$, sets the minimal frequency above which the small-parameter description applies and solid-like modes propagate.

It has taken a long time to verify this picture experimentally. The experimental evidence supporting the propagation of high-frequency modes in liquids currently includes inelastic X-ray, neutron and Brillouin scattering experiments but most important evidence is recent and follows the deployment of powerful synchrotron sources of X-rays \cite{copley,pilgrim,burkel,pilgrim2,ruocco,water,rec-review,hoso,hoso3,mon-na,mon-ga,sn,disu1,disu2,grim,scarponi,water-fast,water-tran}.

Early experiments detected the presence of high-frequency longitudinal acoustic propagating modes and mapped dispersion curves which were in striking resemblance to those in solids \cite{copley}. These and similar results were generated at temperature just above the melting. The measurements were later extended to high temperatures considerably above the melting point, confirming the same result. It is now well established that liquids sustain propagating modes with wavelengths extending down towards interatomic separations, comparable to the wave vectors of phonons in crystals at the Brillouin zone boundaries \cite{pilgrim,burkel,pilgrim2,ruocco,water,rec-review,hoso,hoso3,mon-na,mon-ga,sn,disu1,disu2}. More recently, the same result has been asserted for supercritical fluids \cite{water,disu1,disu2}. Importantly, the propagating modes in liquids include acoustic transverse modes. These were first seen in highly viscous fluids (see, e.g., Refs. \cite{grim,scarponi}), but were then studied in low-viscosity liquids on the basis of positive dispersion \cite{pilgrim,burkel,pilgrim2,rec-review} (the presence of high-frequency transverse modes increases sound velocity from the hydrodynamic to the solid-like value). These studies included water \cite{water-fast}, where it was found that the onset of transverse excitations coincides with the inverse of liquid relaxation time \cite{water-tran}, as predicted by Frenkel \cite{frenkel}.

More recently, high-frequency transverse modes in liquids were directly measured in the form of distinct dispersion branches and verified on the basis of computer modeling \cite{hoso,mon-na,mon-ga,sn,hoso3}, and the striking similarity between dispersion curves in liquids and their crystalline (poly-crystalline) counterparts was noted. We note that the contribution of high-frequency modes is particularly important for liquid thermodynamics because these modes make the largest contribution to the energy due to quadratic density of states.

The above discussion calls for an important question about liquid thermodynamics. In solids, collective modes, phonons, play a central role in the theory, including the theory of thermodynamic properties. Can collective modes in liquids play the same role, in view of the earlier Frenkel proposal and recent experimental evidence? We have started exploring this question \cite{prb} just before the high-frequency transverse modes were directly measured and subsequently developed it in a number of ways \cite{ropp}. This involves calculating the liquid energy as the phonon energy where transverse modes propagate above $\omega_{\rm F}$ in Eq. (\ref{omega}).

The main aim of this paper is to provide direct computational evidence to the phonon theory of liquid thermodynamics and its predictions. We achieve this by calculating the liquid energy and $\omega_{\rm F}$ in extensive molecular dynamics simulations. In the next chapter, we briefly discuss the main steps involved in calculating the liquid energy. We then proceed to calculating the liquid energy and Frenkel frequency independently from molecular dynamics simulations using several methods which agree with each other. We do this for three systems chosen from different classes of liquids: noble, metallic and molecular, and find good agreement between predicted and calculated results in the wide range of temperature and pressure. The range includes both subcritical liquids and supercritical state below the Frenkel line where transverse waves propagate. We calculate and analyze liquid energy and $c_v$ using several different methods. Finally, we discuss how our results offer insights into inter-relationships between structure, dynamics and thermodynamics in liquids and supercritical fluids.

\section{Phonon approach to liquid thermodynamics}

\subsection{Calculating liquid energy}

We summarize the main result of calculation of the liquid energy on the basis of propagating modes. A detailed discussion can be found in a recent review \cite{ropp}.

According to the previous discussion, the propagating modes in liquids include two transverse modes propagating in the solid-like elastic regime with frequency $\omega>\omega_{\rm F}$. The energy of these modes, together with the energy of the longitudinal mode gives the liquid vibrational energy. In addition to vibrations, particles in the liquids undergo diffusive jumps between quasi-equilibrium positions as discussed above. Adding the energy of these jumps to the phonon energy in the Debye model gives the total energy of thermal motion in the liquid \cite{ropp,prb}:

\begin{equation}
E_{\rm T}=NT\left(3-\left(\frac{\omega_{\rm F}}{\omega_{\rm D}}\right)^3\right)
\label{harmo}
\end{equation}

\noindent where $N$ is the number of particles and $\omega_{\rm D}$ is transverse Debye frequency and the subscript refers to thermal motion. Here and below, $k_{\rm B}=1$.

At low temperature, $\tau\gg\tau_{\rm D}$, where $\tau_{\rm D}$ is the Debye vibration period, or $\omega_{\rm F}\ll\omega_{\rm D}$. In this case, Eq. (\ref{harmo}) gives the specific heat $c_v=\frac{1}{N}\frac{dE}{dT}$ close to 3, the solid-like result. At high temperature when $\tau\rightarrow\tau_{\rm D}$ and $\omega_{\rm F}\rightarrow\omega_{\rm D}$, Eq. (\ref{harmo}) gives $c_v$ close to 2. The decrease of $c_v$ from 3 to 2 with temperature is consistent with experimental results in monatomic liquids \cite{grimvall,wallace}. The decrease of $c_v$ is also seen in complex liquids \cite{dexter}.

Eq. (\ref{harmo}) attributes the experimental decrease of $c_v$ with temperature to the reduction of the number of transverse modes above the frequency $\omega_{\rm F}=\frac{1}{\tau}$. The comparison of this effect with experiments can be more detailed if $c_v$ is compared in the entire temperature range where it decreases from $3$ to $2$. This meets the challenge that $\omega_{\rm F}$ in Eq. (\ref{harmo}) is not directly available in the cases of interest. $\omega_{\rm F}$ ($\tau$) is measured is dielectric relaxation or NMR experiments in systems responding to electric or magnetic fields only. These liquids are often complex and do not include simple model systems that are widely studied theoretically such as liquid Ar. Importantly, the range of measured $\omega_{\rm F}$ does not extend to high frequency comparable to $\omega_{\rm D}$, and it is in this range where liquid $c_v$ undergoes an important change from 3 to 2 as discussed above. $\omega_{\rm F}$ can be calculated from the Maxwell relationship $\omega_{\rm F}=\frac{G_\infty}{\eta}$, where $G_\infty$ is the instantaneous shear modulus and $\eta$ is viscosity taken from a different experiment \cite{ropp}. More recently, it has been suggested \cite{puosi} that taking the shear modulus at a finite high frequency (rather than infinite frequency) agrees better with the modelling data. Apart from rare estimations \cite{puosi,wallace-G}, $G_\infty$ is not available. In practice, the comparison of experimental $c_v$ and $c_v$ predicted as $\frac{dE}{dT}$ with $E$ given by Eq. (\ref{harmo}) is done by keeping $G_\infty$ as a free parameter, obtaining a good agreement between experimental and predicted $c_v$ and observing that $G_\infty$ lies in the range of several GPa typical for liquids \cite{prb,ropp}. In the last few years, Eq. (\ref{harmo}) and its extensions to include the phonon anharmonicity and quantum effects of phonon excitations was shown to account for the experimental $c_v$ of over 20 different systems, including metallic, noble, molecular and network liquids \cite{ropp}.

In view of the persisting problem of liquid thermodynamics, it is important to test Eq. (\ref{harmo}) directly by linking the liquid energy ($c_v$) on one hand and $\omega_{\rm F}$ on the other and testing the theory in a precise way. This, together with achieving consistency with other approaches to calculate the liquid energy, is one of the objectives of this study. Importantly, this programme includes supercritical fluids as well as subcritical liquids, as discussed below.

\subsection{Thermodynamics of supercritical fluids}

\begin{figure}
\begin{center}
{\scalebox{0.45}{\includegraphics{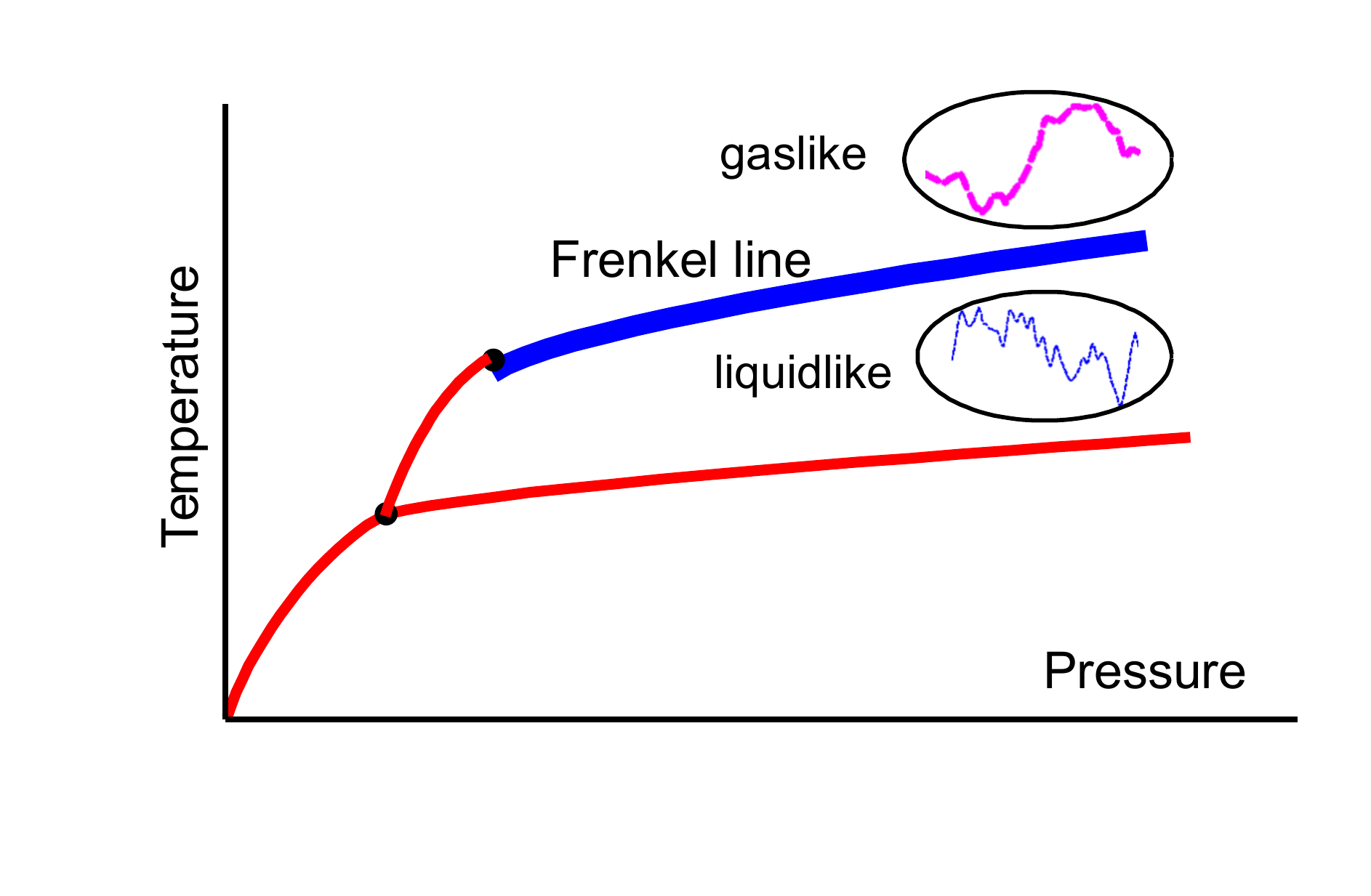}}}
\end{center}
\caption{Colour online. The Frenkel line in the supercritical region. Particle dynamics includes both oscillatory and diffusive components below the line, and is purely diffusive above the line. Below the line, the system is able to support rigidity and transverse modes at high frequency. Above the line, particle motion is purely diffusive, and the ability to support rigidity and transverse modes is lost at all available frequencies. Crossing the Frenkel line from below corresponds to the transition between the ``rigid'' liquid to the ``non-rigid'' gas-like fluid.}
\label{frenline}
\end{figure}

If the system is below the critical point (see Figure 1), the temperature increase eventually results in boiling and the first-order transition, with $c_v$ discontinuously decreasing to about $\frac{3}{2}$ in the gas phase. The intervening phase transition excludes the state of the liquid where $c_v$ can gradually reduce to $\frac{3}{2}$ and where interesting physics operates. However, this becomes possible above the critical point. This brings us to the interesting discussion of the supercritical state of matter. Theoretically, little is known about the supercritical state, apart from the general assertion that supercritical fluids can be thought of as high-density gases or high-temperature fluids whose properties change smoothly with temperature or pressure and without qualitative changes of properties. This assertion followed from the known absence of a phase transition above the critical point. We have recently proposed that this picture should be modified, and that a new line, the Frenkel line (FL), exists above the critical point and separates two states with distinct properties (see Figure \ref{frenline}) \cite{pre,prl,phystoday, ufn}. Physically, the FL is not related to the critical point and exists in systems where the critical point is absent.

The main idea of the FL lies in considering how the particle dynamics change in response to pressure and temperature. Recall that particle dynamics in the liquid can be separated into solid-like oscillatory and gas-like diffusive components. This separation applies equally to supercritical fluids as it does to subcritical liquids. Indeed, increasing temperature reduces $\tau$, and each particle spends less time oscillating and more time jumping; increasing pressure reverses this and results in the increase of time spent oscillating relative to jumping. Increasing temperature at constant pressure or density (or decreasing pressure at constant temperature) eventually results in the disappearance of the solid-like oscillatory motion of particles; all that remains is the diffusive gas-like motion. This disappearance represents the qualitative change in particle dynamics and gives the point on the FL in Figure \ref{frenline}. Most important system properties qualitatively change either on the line or in its vicinity \cite{pre,prl,phystoday,ufn}. In a given system, the FL exists at arbitrarily high pressure and temperature, as does the melting line.

Quantitatively, the FL can be rigorously defined by pressure and temperature at which the minimum of the velocity autocorrelation function (VAF) disappears \cite{prl}. Above the line defined in such a way, velocities of a large number of particles stop changing their sign and particles lose the oscillatory component of motion. Above the line, VAF is monotonically decaying as in a gas \cite{prl}. For the purposes of this discussion, the significance of the FL is that the phonon approach to liquids and Eq. (\ref{harmo}) apply to supercritical fluids below the FL to the same extent as they apply to subcritical liquids. Indeed, the presence of an oscillatory component of particle motion below the FL implies that $\tau$ is a well-defined parameter and that transverse modes propagate according to Eq. (\ref{omega}). The ability of the supercritical system to sustain solid-like rigidity at frequency above $\omega_{\rm F}$ suggested the term ``rigid'' liquid to differentiate it from the ``non-rigid'' gas-like fluid above the FL \cite{pre,prl}.

Therefore, the FL separates the supercritical state into two states where transverse modes can and cannot propagate. This is supported by direct calculation of the current correlation functions \cite{condmat} showing that propagating and non-propagating transverse modes are separated by the Frenkel line. Interestingly, Eq. (\ref{harmo}) can serve as a thermodynamic definition of the FL: the loss of the oscillatory component of particle motion at the FL approximately corresponds to $\tau\rightarrow\tau_{\rm D}$ (here, $\tau_{\rm D}$ refers to Debye period of transverse modes) or $\omega_{\rm F}\rightarrow\omega_{\rm D}$. According to Eq. (\ref{harmo}), this gives $c_v$ of about 2. Using the criterion $c_v=2$ gives the line that is in remarkably good coincidence with the line obtained from the VAF criterion above \cite{prl}.

\section{Simulation details}

We have considered liquids from three important system types: noble Ar, molecular CO$_2$ and metallic Fe. We have used the molecular dynamics (MD) simulation package DL\_POLY \cite{dlpoly} and simulated systems with $4576 - 8000$ particles with periodic boundary conditions. The interatomic potential for Ar is the pair Lennard-Jones potential \cite{ar}, known to perform well at elevated pressure and temperature. For CO$_2$ and Fe, we have used interatomic potentials optimized tested in the liquid state at high pressure and temperature. The potential for CO$_2$ is the rigid-body nonpolarizable potential based on a quantum chemistry calculation, with the partial charges derived using the Distributed Multipole Analysis method \cite{min}. Fe was simulated using the many-body embedded-atom potential \cite{iron}. In the case of CO$_2$, the electrostatic interactions were evaluated using the smooth particle mesh Ewald method. The MD systems were first equilibrated in the constant pressure and temperature ensemble at respective state points for 20 ps. System properties were subsequently simulated at different temperatures and averaged in the constant energy and volume ensemble for 30 ps.

We are interested in properties of real dense strongly-interacting liquids with potential energy comparable to kinetic energy and hence have chosen fairly high densities: $\rho=1.5$ g/cm$^3$ and $\rho=1.9$ g/cm$^3$ for Ar, $\rho=8$ g/cm$^3$ and $\rho=11$ g/cm$^3$ for Fe and $\rho=1.34$ g/cm$^3$ for CO$_2$. The lowest temperature in each simulation was the melting temperature at the corresponding density, $T_m$. The highest temperature significantly exceeded the temperature at the Frenkel line at the corresponding density,
$T_{\rm F}$, taken from the earlier calculation of the Frenkel line in Ar \cite{prl}, Fe \cite{scirep} and CO$_2$ \cite{yang}. As discussed above, the temperature range between $T_m$ and $T_{\rm F}$ corresponds to the regime where transverse modes progressively disappear and where Eq. (\ref{harmo}) applies. We have simulated $100-700$ temperature points at each pressure depending on the system. The number of temperature points was chosen to keep the temperature step close to 10 K.

As discussed above, Eq. (\ref{harmo}) applies to subcritical liquids as well as to supercritical fluids below the Frenkel line. Accordingly, our simulations include the temperature range both below and above the critical temperature. This will be discussed in more detail below.

\section{Results and discussion}

\subsection{Liquid energy and heat capacity}

We have calculated $\omega_{\rm F}$ in (\ref{harmo}) from its definition in (\ref{omega}), as $\omega_{\rm F}=\frac{1}{\tau}$. $\tau$ can be calculated in a number of ways. Most common methods calculate $\tau$ as decay time of the self-intermediate scattering or other functions by the factor of $e$ or as the time at which the mean-squared displacement crosses over from ballistic to diffusive regime \cite{overlap}. These methods give $\tau$ in agreement with a method employing the overlap function depending on the cutoff parameter $a_c$ provided $a_c=0.3a$, where $a$ is the inter-molecuar distance \cite{overlap}. We use the latter method and calculate $\tau$ at 13-20 temperature points at each density depending on the system. At each density, we fit $\tau$ to the commonly used Vogel-Fulcher-Tammann dependence and use $\omega_{\rm F}=\frac{1}{\tau}$ to calculate the liquid energy predicted from the theory. The predicted $c_v$ is calculated as $c_v=\frac{1}{N}\frac{dE}{dT}$ where $E$ is given by Eq. (\ref{harmo}):

\begin{equation}
c_v=3-\left(\frac{\omega_{\rm F}}{\omega_{\rm D}}\right)^3-\frac{3T\omega_{\rm F}^2}{\omega_{\rm D}^3}\frac{d\omega_{\rm F}}{dT}
\label{cv}
\end{equation}

\noindent where $N$ is the number of atoms for Ar and Fe and the number of molecules for CO$_2$.

The first two terms in (\ref{cv}) give $c_v=2$ when $\omega_{\rm F}$ tends to its high-temperature limit of $\omega_{\rm F}$. The last term reduces $c_v$ below 2 by a small amount because $\frac{d\omega_{\rm F}}{dT}$ is close to zero at high temperature \cite{ropp}.

We now compare the calculated energy and $c_v$ with those directly computed in the MD simulations. We note that the energy in Eq. (\ref{harmo}) is the energy of thermal phonon motion, $E_T$, which contributes to the total liquid energy as

\begin{equation}
E=E_0+E_{\rm T}
\label{toten}
\end{equation}

\noindent where $E_0$ is liquid energy at zero temperature and represents temperature-independent background contribution due to the interaction energy.

\begin{figure}
\begin{center}
{\scalebox{0.65}{\includegraphics{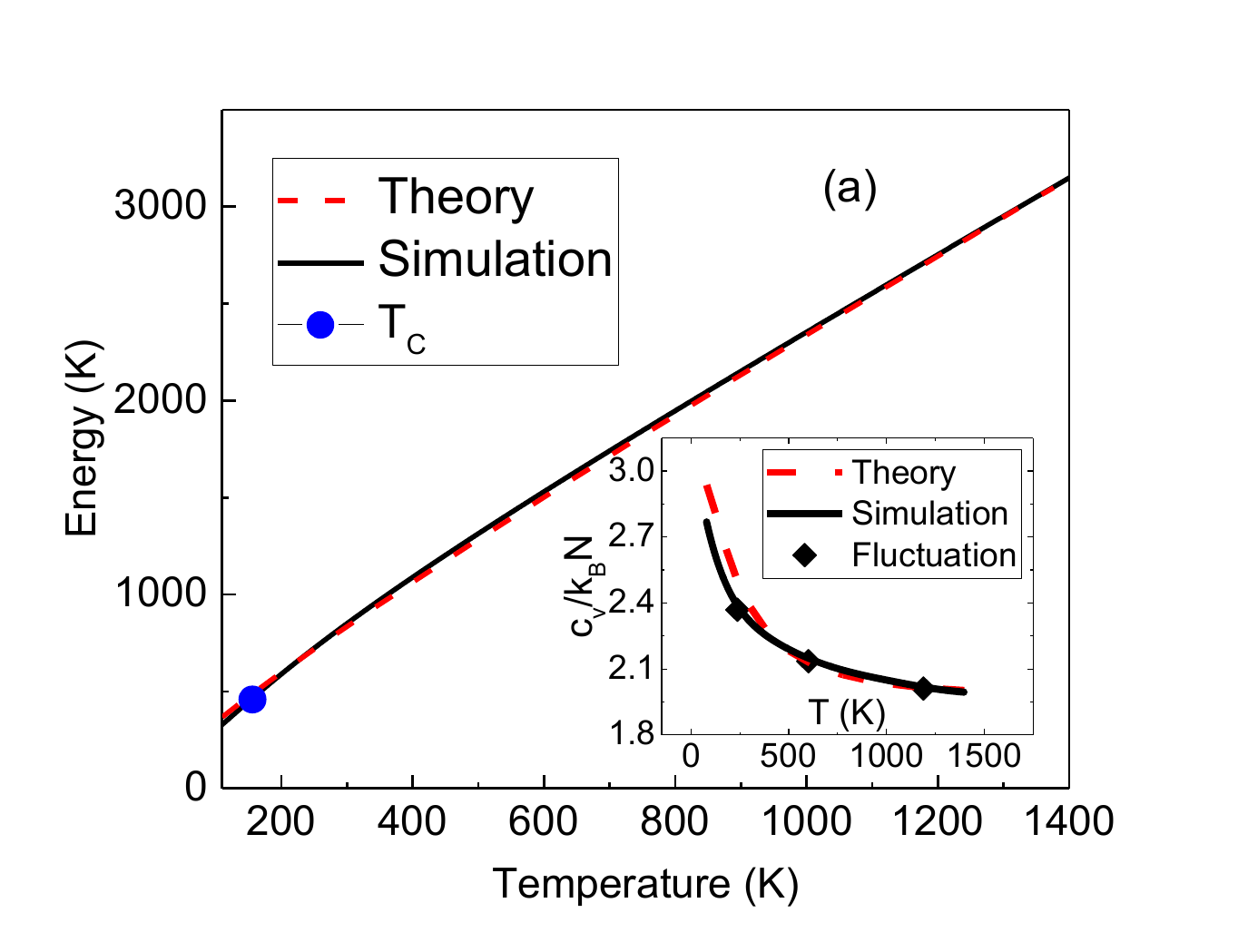}}}
{\scalebox{0.65}{\includegraphics{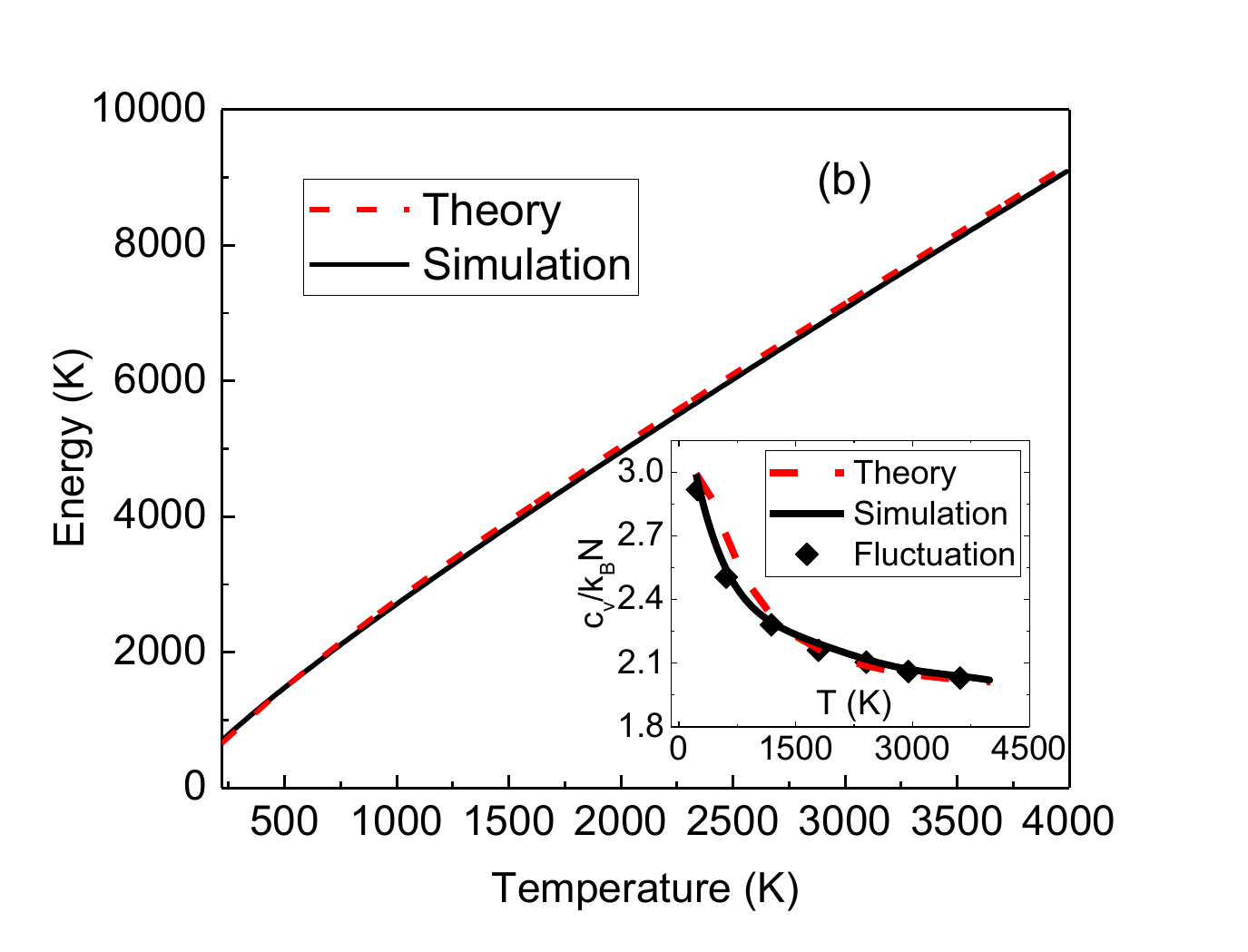}}}
\end{center}
\caption{Energy per particle and specific heat of Ar at density $\rho=1.5$ g/cm$^3$ (a) and $\rho=1.9$ g/cm$^3$ (b). Solid and dashed lines correspond to results from simulations and theory, respectively. The large (blue) circle corresponds to critical temperature. The black solid curves in the insets show $c_v$ calculated as $c_v=\frac{1}{N}\frac{dE}{dT}$. Solid diamonds correspond to $c_v$ calculated from the fluctuation formula (see text). The red (dashed) line is the theoretical result for $c_v$.}
\label{ar}
\end{figure}

In comparing the calculated $E_{\rm T}$ in Eq. (\ref{harmo}) with the energy from MD simulations, we therefore subtract the constant term from the MD energy. The comparison of $c_v=\frac{1}{N}\frac{dE}{dT}$ is performed directly because the constant term does not contribute to $c_v$. We have also calculated $c_v$ using the fluctuations formula for the kinetic energy $K$ in the constant energy ensemble: $\langle K^2\rangle-\langle K\rangle^2=1.5(k_{\rm B} T)^2N(1-1.5k_{\rm B}/c_v)$ \cite{frensim}. Both methods agree well, as follows from Figures \ref{ar}a and \ref{ar}b.

\begin{figure}
\begin{center}
{\scalebox{0.65}{\includegraphics{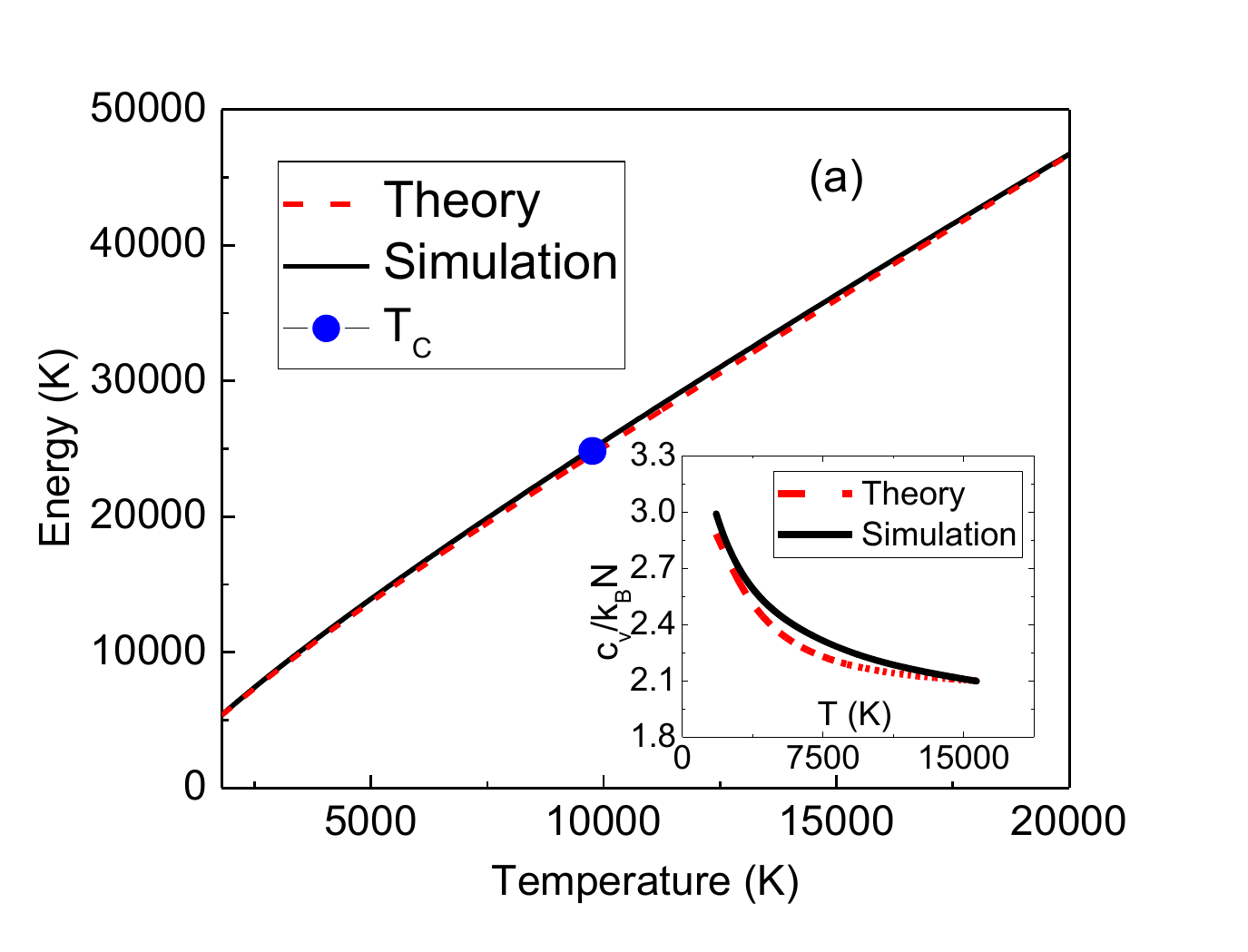}}}
{\scalebox{0.65}{\includegraphics{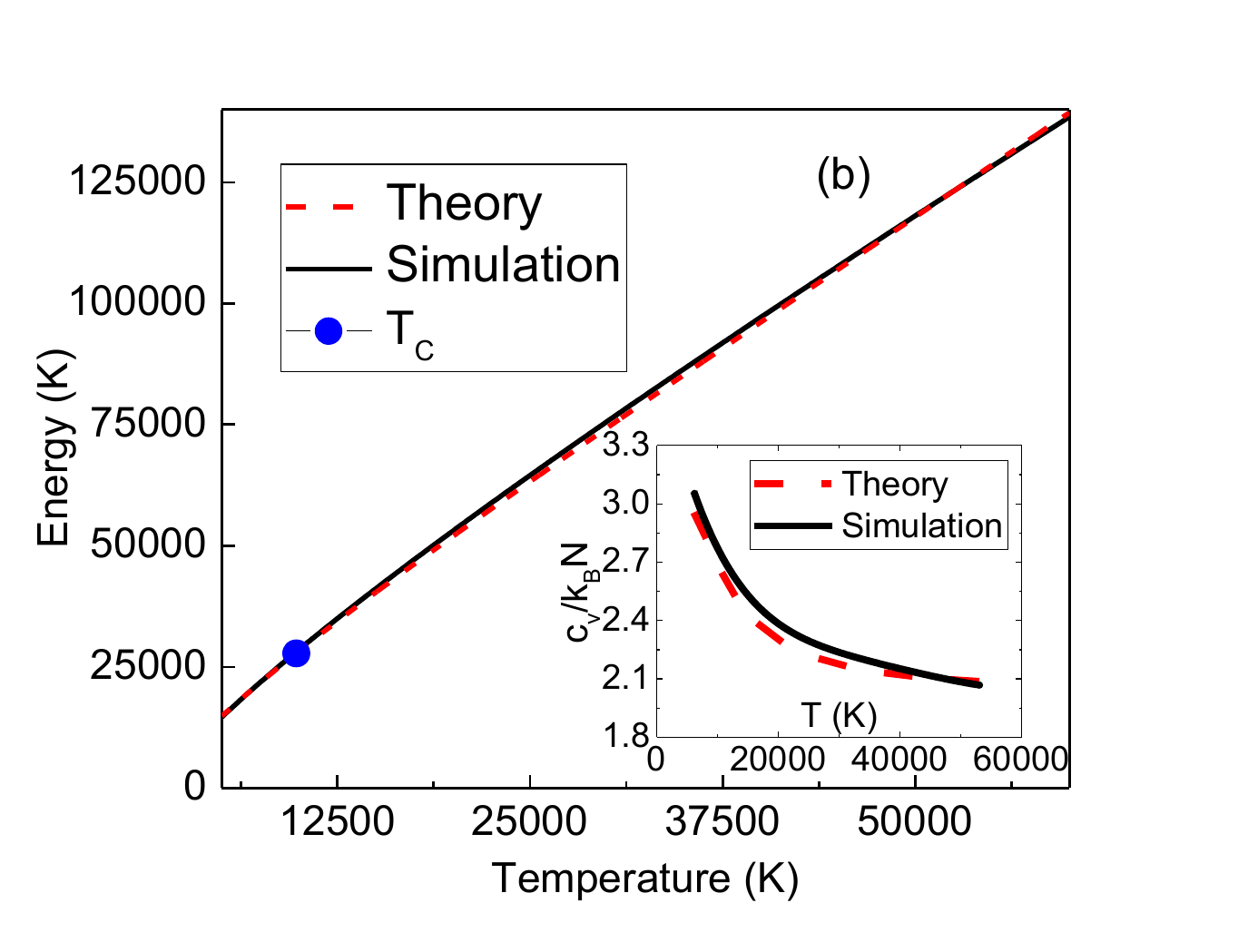}}}
\end{center}
\caption{Energy per particle and specific heat of Fe at density $\rho=8$ g/cm$^3$ (a) and $\rho=11$ g/cm$^3$ (b). Solid and dashed lines correspond to results from simulations and theory, respectively. The large (blue) circle corresponds to critical temperature. The black solid curves in the insets show $c_v$ calculated as $c_v=\frac{1}{N}\frac{dE}{dT}$. The red (dashed) line is the theoretical result for $c_v$.}
\label{fe}
\end{figure}

There is only one adjustable parameter in Eq. (\ref{harmo}), $\omega_{\rm D}$, which is expected to be close to transverse Debye frequency. $\omega_{\rm F}$ is independently calculated from the MD simulation as discussed above. In Figures \ref{ar} and \ref{fe} we compare the energy and $c_v$ calculated on the basis of Eqs. (\ref{harmo}) and (\ref{cv}) and compare them with those computed in MD simulations. Blue circle in each figure shows the critical temperature. We observe good agreement between predicted and calculated properties in a temperature range including both subcritical and supercritical temperature. This involved using $\tau_{\rm D}\approx 0.6$ ps ($\rho=8$ g/cm$^3$) and $\tau_{\rm D}\approx 0.2$ ps ($\rho=11$ g/cm$^3$) for Fe, $\tau_{\rm D}\approx 0.9$ ps ($\rho=1.5$ g/cm$^3$) and $\tau_{\rm D}\approx 0.3$ ps ($\rho=1.9$ g/cm$^3$) for Ar and $\tau_{\rm D}\approx 0.5$ ps for CO$_2$, in reasonable order-of-magnitude agreement with experimental $\tau_{\rm D}$ of respective crystalline systems as well as maximal frequencies seen in experimental liquid dispersion curves (see, e.g., \cite{hoso3}). We note the expected trend of $\tau_{\rm D}$ reducing with density.

At high temperature where $\omega_{\rm F}\approx\omega_{\rm D}$, Eq. (\ref{cv}) predicts $c_v$ close to 2, noting that the last term gives only a small contribution to $c_v$ because $\omega_{\rm F}$ becomes slowly varying at high temperature. Consistent with this prediction, we observe the decrease of $c_v$ from 3 to 2 in Figures \ref{ar} and \ref{fe}.

The agreement between the predicted and calculated results supports the interpretation of the decrease of $c_v$ with temperature discussed in the Introduction: $\omega_{\rm F}$ decreases with temperature, and this causes the reduction of the number of transverse modes propagating above $\omega_{\rm F}$ and hence the reduction of $c_v$.

\begin{figure}
\begin{center}
{\scalebox{0.65}{\includegraphics{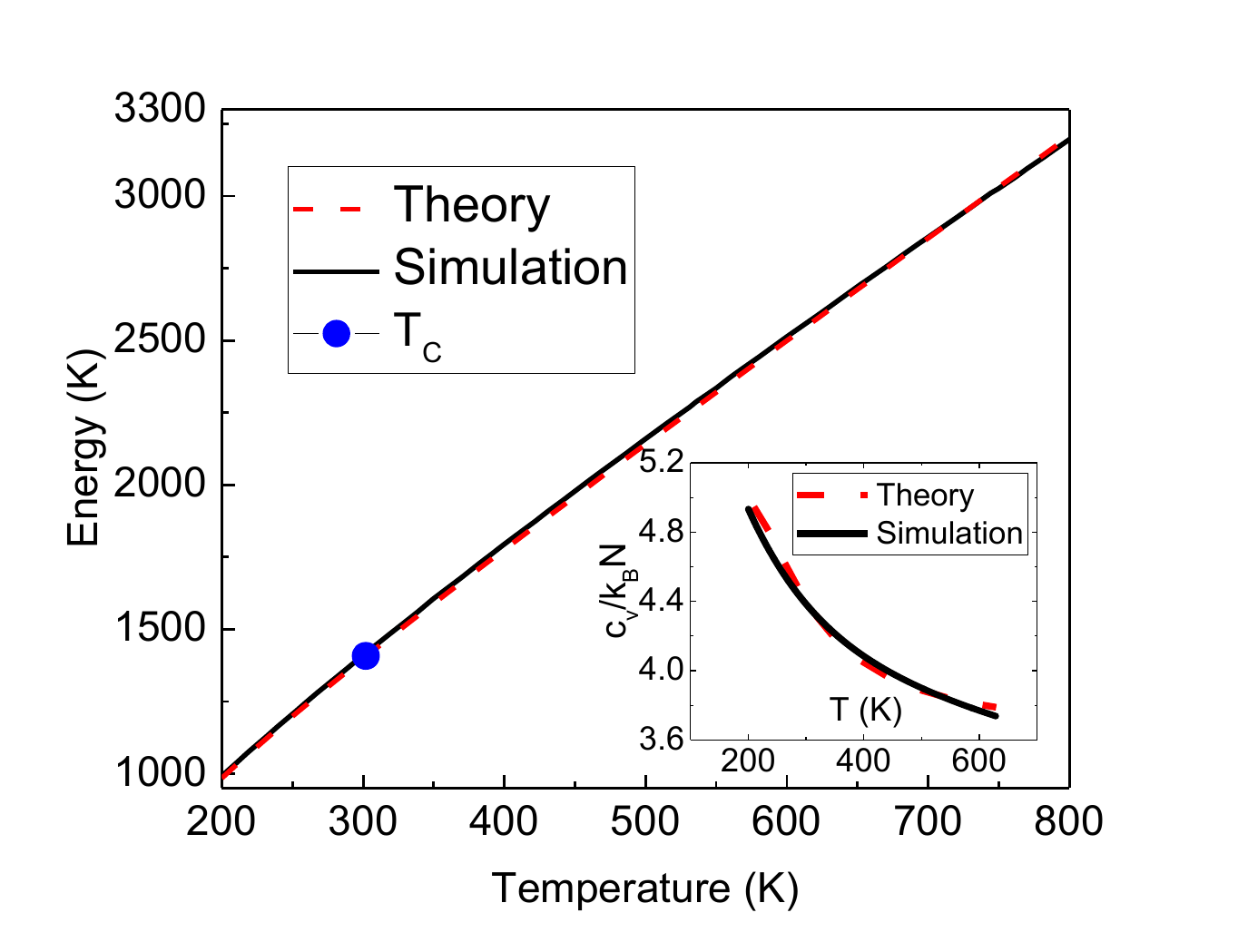}}}
\end{center}
\caption{Energy per particle and specific heat of CO$_2$ at density $\rho=1.34$ g/cm$^3$. Solid and dashed lines correspond to results from simulations and theory, respectively. The large (blue) circle corresponds to critical temperature. The black solid curves in the insets show $c_v$ calculated as $c_v=\frac{1}{N}\frac{dE}{dT}$. The red (dashed) line is the theoretical result for $c_v$.}
\label{co2}
\end{figure}

For CO$_2$, the same mechanism operates except we need to account for degrees of freedom in a molecular system. We first consider the case of solid CO$_2$. The MD interatomic potential treats CO$_2$ molecules as rigid linear units, contributing the kinetic term of 2.5 to the specific heat per molecule including 1 from the rotational degrees of freedom of the linear molecular and 1.5 from translations (here, we have noted that CO$_2$ molecules librate and rotate in the solid at low and high temperature, respectively \cite{rotat}). Noting the potential energy contributes the same term due to equipartition, the specific heat becomes 5 per molecule. This implies that for molecular CO$_2$, Eqs. (\ref{harmo}) modifies as $E_{\rm T}=NT\left(5-\left(\frac{\omega_{\rm F}}{\omega_{\rm D}}\right)^3\right)$, where $N$ is the number of molecules and $\omega_{\rm F}$ is related to the jump frequency of molecules and which gives $c_v=5$ in the solid state where $\omega_{\rm F}$ is infinite. We use the modified equation to calculate the energy and $c_v$ and compare them to those computed from the MD simulation in Figure \ref{co2}.

Consistent with the above discussion, we observe that $c_v$ for CO$_2$ calculated directly from the MD simulations is close to 5 at low temperature just above melting. At this temperature, $\omega_{\rm F}\ll\omega_{\rm D}$, giving the solid-like value of $c_v$ as in the case of monatomic Ar and Fe. As temperature increases, two transverse modes of inter-molecular motion progressively disappear, resulting in the decrease of $c_v$ to the value of about $c_v=4$, in agreement with $c_v$ calculated from the theoretical equation for $E_{\rm T}$.


We note that the temperature range in which we compare the predicted and calculated properties is notably large (e.g., $200-8000$ K for Ar, and $2000-55000$ K for Fe). This range is 10-100 times larger than those typically considered earlier \cite{ropp}. The higher temperatures for Fe might appear as unusual, however we note that liquid iron as well as supercritical iron fluid remains an unmodified system up to very high temperature: the first ionization potential of Fe is 7.9 eV, or over 90,000 K. Hence the considered temperature range is below the temperature at which the system changes its structure and type of interactions. 

The very wide temperature range reported here is mostly related to the large part of the temperature interval in Figures \ref{ar}-\ref{co2} being above the critical point where no phase transition intervenes and where the liquid phase exists at high temperature, in contrast to subcritical liquids where the upper temperature is limited by the boiling line. The agreement between predicted and calculated properties in such a wide temperature range adds support to the phonon approach to liquid thermodynamics we propose.

We make three points regarding the observed agreement between the calculated and predicted results. First, the collective modes contributing to the thermal energy in (\ref{harmo}) are considered to be harmonic. The anharmonicity can be accounted for in the Gr\"{u}neisen approximation, however this involves an additional parameter \cite{ropp}. We attempted to avoid introducing additional parameters and sought to test Eq. (\ref{harmo}) which contains only one parameter, $\omega_{\rm D}$.

Second, Eq. (\ref{harmo}) involves the Debye model and quadratic density of states (DOS). This approximation is justified since the Debye model is particularly relevant for disordered isotropic systems such as glasses \cite{landau}, which are known to be nearly identical to liquids from the structural point of view. Furthermore, the experimental dispersion curves in liquids are very similar to those in solids such as poly-crystals \cite{mon-na,mon-ga,sn}. Therefore, the Debye model can be used in liquids to the same extent as in solids. One important consequence of this is that the high-frequency range of the phonon spectrum makes the largest contribution to the energy, as it does in solids including disordered solids. We also note that liquid DOS can be represented as the sum of solid-like and gas-like components in the two-phase thermodynamic model \cite{goddard}, and the solid-like component can be extracted from the liquid DOS calculated in MD simulations. This can provide more information about the DOS beyond Debye approximation.

Third, Eq. (\ref{harmo}) assumes a lower frequency cutoff for transverse waves, $\omega_{\rm F}=\frac{1}{\tau}$, as envisaged by Frenkel in (\ref{omega}). Our recent detailed analysis of the Frenkel equations shows that the dispersion relationship for liquid transverse modes is $\omega=\sqrt{c_s^2k^2-\frac{1}{4\tau^2}}$, where $c_s$ is the shear speed of sound and $k$ is wavenumber \cite{ropp}. Here, $\omega$ gradually crosses over from $0$ to its solid-like branch $\omega=c_sk$ when $\omega\gg\omega_{\rm F}=\frac{1}{\tau}$. In this sense, using a lower frequency cutoff in (\ref{harmo}) might be thought of as an approximation. However, we have recently shown \cite{chenxing} that the square-root dependence of $\omega$ gives the liquid energy that is identical to (\ref{harmo}).

\subsection{Structural crossover and its relationship to dynamical and thermodynamic properties}

The results in the previous sections support the picture in which the decrease of liquid $c_v$ from 3 to 2 is related to reduction of the energy of transverse modes propagating above $\omega_{\rm F}$ as described by Eq. (\ref{cv}). According to Eq. (\ref{cv}), $c_v=2$ corresponds to complete disappearance of transverse modes at the FL when $\omega_{\rm F}\approx\omega_{\rm D}$ (the disappearance is supported by the direct calculation of transverse modes on the basis of current correlation functions \cite{condmat}). Importantly, $c_v=2$ marks the crossover of $c_v$ because the evolution of collective modes is qualitatively different below and above the FL \cite{ropp}. Below the line, transverse modes disappear starting from the lowest frequency $\omega_{\rm F}$. Above the line, the remaining longitudinal mode starts disappearing starting from the highest frequency $\frac{2\pi c}{L}$, where $L$ is the particle mean free path (no oscillations can take place at distance smaller than $L$). This gives qualitatively different behavior of the energy and $c_v$ below and above the FL, resulting in their crossover at the FL \cite{ropp}.

Interestingly, the thermodynamic crossover at $c_v=2$ implies a structural crossover. Indeed, the energy per particle in a system with pair-wise interactions is

\begin{equation}
E=\frac{3}{2}k_{\rm B}T+4\pi \rho \int\limits_0^{\infty} r^2 U(r)g(r)dr
\label{ene}
\end{equation}
\noindent where $\rho=N/V$ is number density and $g(r)$ is radial distribution function.

According to Eq. (\ref{toten}), the liquid energy is $E=E_0+E_{\rm T}$, where $E_{\rm T}$ is given by Eq. (\ref{harmo}). If the system energy undergoes the crossover at the FL where $c_v=2$, Eq. (\ref{ene}) implies that $g(r)$ should also undergo a crossover. Therefore, the structural crossover in liquids can be predicted on the basis of the thermodynamic properties.

\begin{figure}
\begin{center}
{\scalebox{0.6}{\includegraphics{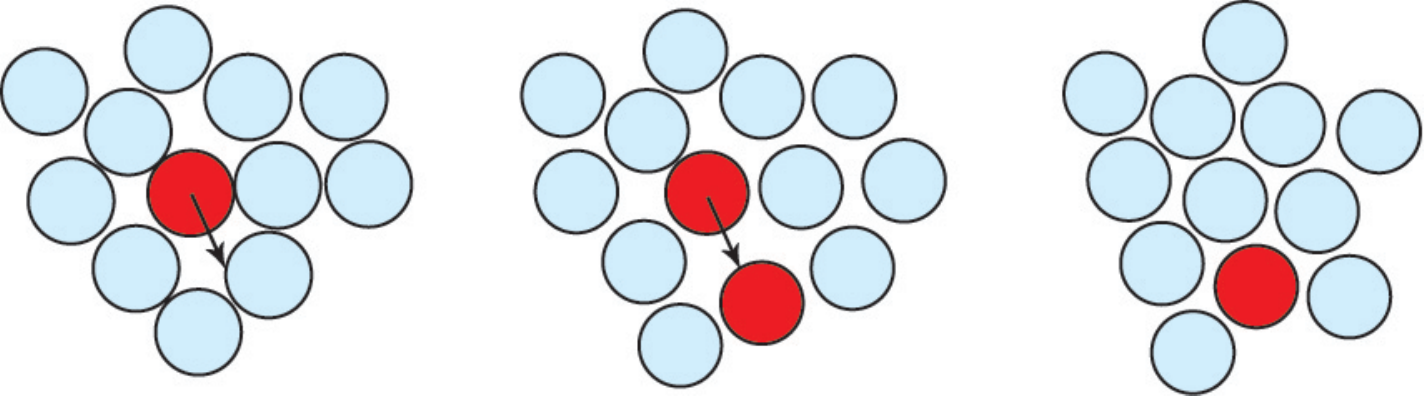}}}
\end{center}
\caption{Schematic representation of a jump event in the liquid.}
\label{atoms}
\end{figure}

We also expect the structural crossover at the FL to be related to the dynamical crossover on general grounds. As discussed above, below the FL particles oscillate around quasi-equilibrium positions and occasionally jump between them. The average time between jumps is given by liquid relaxation time, $\tau$ (Figure \ref{atoms} schematically shows a local jump event from its surrounding ``cage''.) This means that a static structure exists during $\tau$ for a large number of particles below the FL, giving rise to the well-defined medium-range order comparable to that existing in disordered solids \cite{salmon}. On the other hand, the particles lose the oscillatory component of motion above the FL and start to move in a purely diffusive manner as in gases. This implies that the features of $g(r)$ are expected to be gas-like. As a result, $g(r)$ medium-range peaks are expected to have different temperature dependence below and above the FL. This behavior was observed in Ar in MD simulations in the short-range structure \cite{jcp}. More recently, the crossover in supercritical Ne in the medium range at the FL was ascertained on the basis of X-ray scattering experiments \cite{clemens}.

\begin{figure}
\begin{center}
{\scalebox{0.7}{\includegraphics{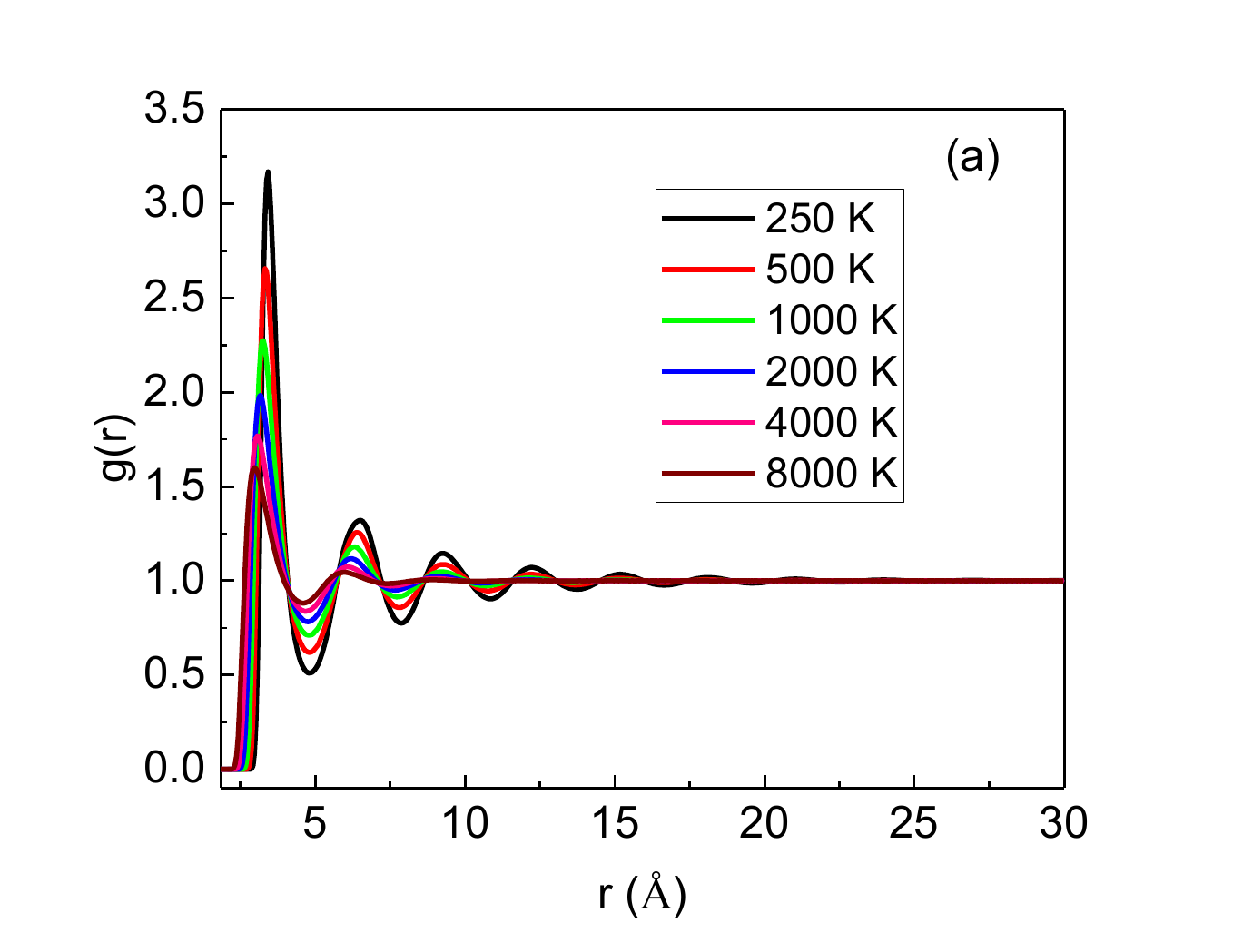}}}
{\scalebox{0.7}{\includegraphics{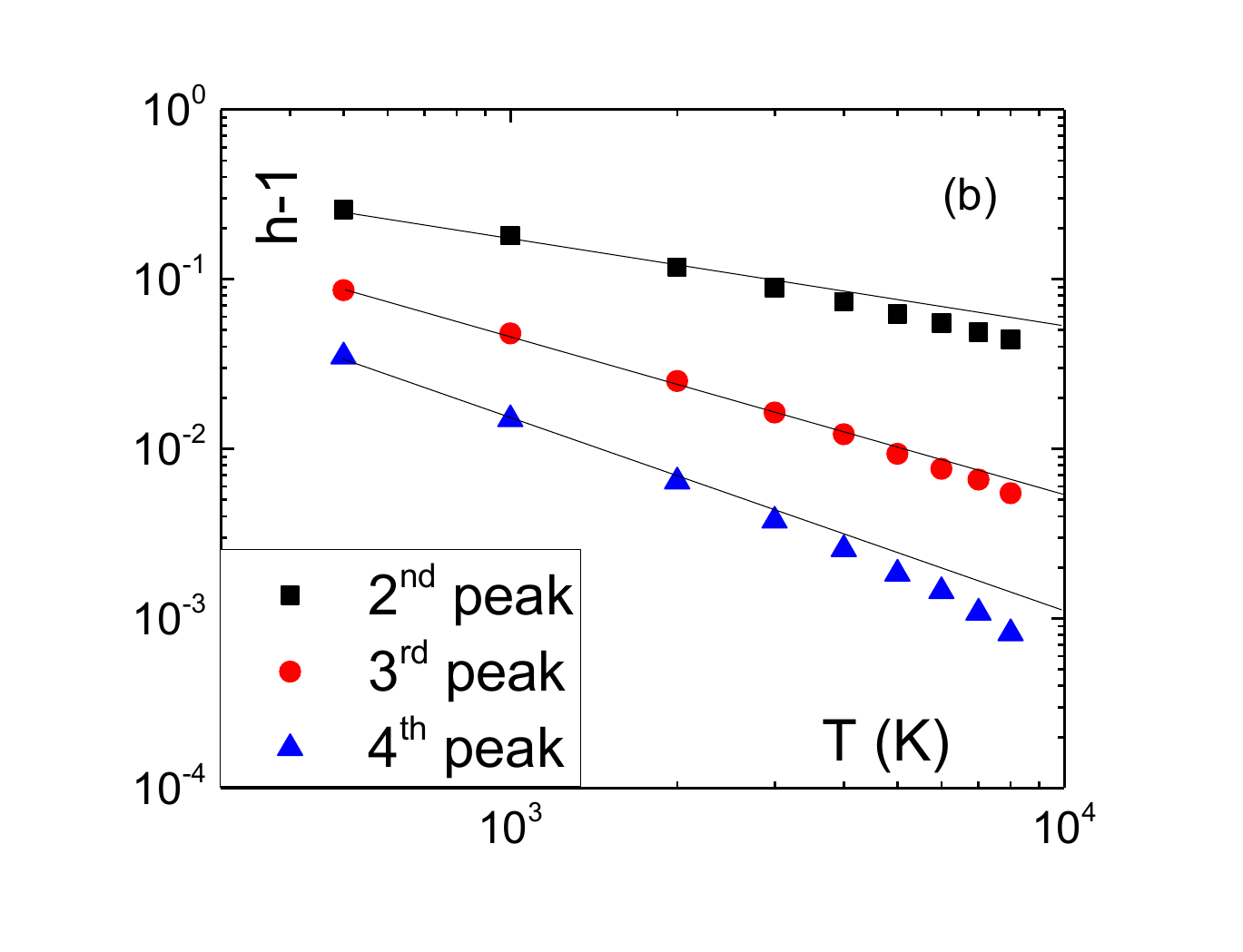}}}
\end{center}
\caption{(a) Pair distribution functions of Ar at different temperatures. The temperatures correspond to the first peak decreasing from top to bottom at 250 K, 500 K, 1000 K, 2000 K, 4000 K and 8000 K; (b) $h-1$, where $h$ are the heights of PDF peaks at different temperatures. The lines are linear fits to the low-temperature data range.}
\label{rdf}
\end{figure}

In Figure \ref{rdf}a we plot pair distribution functions (PDFs) of Ar at density $\rho=1.9$ g/cm$^3$ in a wide temperature range. Using the FL criterion $c_v=2$ gives the temperature at the FL, $T_{\rm F}$, of about 4000 K at that density, which we find to be consistent with the criterion of the disappearance of the minimum of the velocity autocorrelation function \cite{prl}. The PDF was calculated with the distance step of $0.05$ \AA, giving 600 PDF points.

We observe PDF peaks in the medium range order up to about 20 \AA\ at low temperature. The peaks reduce and broaden with temperature. To study this in more detail, we plot the peak heights vs temperature in Figure \ref{rdf}b. We observe that the medium-range third and fourth peaks persist well above the critical temperature ($T_c=151$ K for Ar): the highest temperature simulated corresponds to $53T_c$. This interestingly differs from the traditional expectation that the structure of the matter so deep in the supercritical state has gas-like features only. At temperature above $T_{\rm F}$, the height of the fourth peak becomes comparable to its temperature fluctuations (calculated as the standard deviation of the peak height over many structures separated in time by 1 ps at each temperature) by order of magnitude. The fifth and higher-order peaks disappear before the highest temperature in the simulated range is reached.

We plot the peak heights in Figure \ref{rdf}b in the double-logarithmic plot because we expect to see an approximate power-law decay of the peak heights at low temperature. Indeed, PDF in solids can be represented as a set of Gaussian functions with peaks heights $h$ depending on temperature as $h\propto\frac{1}{\sqrt{T}}\exp\left(-\frac{\alpha}{T}\right)$ where $\alpha$ is a temperature-independent factor \cite{frenkel,marad}. This temperature dependence of $h$ was also quantified in MD simulations \cite{stan}. $h$ decrease mostly due to the factor $\frac{1}{\sqrt{T}}$ whereas the effect of the exponential factor on $h$ is small and serves to reduce the rate at which $h$ decrease \cite{stan}. This implies that in solids, $\log h\propto-\log T$ approximately holds.

In liquids, we expect the same relationship to hold below the FL where $\tau\gg\tau_{\rm D}$, corresponding to a particle oscillating many times before diffusively moving to the next quasi-equilibrium position. Indeed, the ratio of the number of diffusing particles $N_{\rm dif}$ to the total number of particles $N$ in the equilibrium state is $\frac{N_{\rm dif}}{N}=\frac{\tau_{\rm D}}{\tau}$ \cite{ropp} at any given moment of time. $\frac{N_{\rm dif}}{N}$ is small when $\tau\gg\tau_{\rm D}$ below the FL and can be neglected. Hence, $\log h\propto-\log T$ applies to liquids at any given moment of time below the FL where $\tau\gg\tau_{\rm D}$. This also applies to longer observation times if $h$ is averaged over $\tau$ \cite{ropp}. We note that the above result, $h\propto\frac{1}{\sqrt{T}}$, involves the assumption that the energy of particle displacements is harmonic (see, e.g., Ref. \cite{frenkel}). Anharmonicity becomes appreciable at high temperature, however the anharmonic energy terms are generally small compared to the harmonic energy. This is witnessed by the closeness of high-temperature $c_v$ to its harmonic result for both solids and high-temperature liquids \cite{andr,bolm}.

We therefore expect that $\log(h-1)\propto-\log T$ approximately holds in the low-temperature range below the FL as in solids but deviates from the linearity around the crossover at the FL where $\tau\rightarrow\tau_{\rm D}$ and where the dynamics becomes gas-like (the calculated PDF in Fig. \ref{rdf}a is normalized to 1 where no correlations are present at large distances; hence we plot $h-1$ in order to compare it with the theoretical result $h\propto\frac{1}{\sqrt{T}}$ which tends to zero when no correlations are present at high temperature). We note that the crossover is expected to be broad because $\tau\gg\tau_{\rm D}$ applies well below the FL only. A substantial diffusive motion takes place in the vicinity of the line where $\frac{N_{\rm dif}}{N}$ can not be neglected, affecting the linear relationship.

Consistent with the above prediction, we observe the linear regime at low temperature in Figure \ref{rdf}b, followed by the deviation from the straight lines taking place around 3000 K for the 2nd peak, 5000 K for the 3rd peak and 4000 K for the 4th peak, respectively. The smooth crossover in the 3000-5000 K range is centered around 4000 K, consistent with the temperature at the Frenkel line discussed above. We also note that 4000 K corresponds to the specific heat $c_v=2$ in Figure \ref{ar}b, in agreement with the earlier discussion.

\section{Summary}

As discussed in the Introduction, liquids have been viewed as inherently complicated systems lacking useful theoretical concepts such as a small parameter \cite{landau}. Together with recent experimental evidence and theory \cite{ropp}, the modelling data presented here and its quantitative agreement with predictions are beginning to change this traditional perspective. Our extensive molecular dynamics simulations of liquid energy and specific heat provide direct evidence for the link between dynamical and thermodynamic properties of liquids. We have found this to be the case for several important types of liquids at both subcritical and supercritical conditions spanning thousands of Kelvin. This supports an emerging picture that liquid thermodynamics can be understood on the basis of high-frequency collective modes. A more general implication is that, contrary to the prevailing view, liquids are emerging as systems amenable to theoretical understanding in a consistent picture as is the case in solid state theory. In addition to the link between dynamical and thermodynamic properties, we have discussed how these properties are related to liquid structure.

This research utilised MidPlus computational facilities supported by QMUL Research-IT and funded by the EPSRC Grant EP/K000128/1. We acknowledge the support of the Royal Society, RFBR (15-52-10003) and CSC.

\end{document}